# Coverage, field specialization and impact of scientific publishers indexed in the 'Book Citation Index'

Daniel Torres-Salinas[1], Nicolás Robinson-García[2]*, J.M. Campanario[3] and Emilio Delgado López-Cózar[2]

[1] EC3: Evaluación de la Ciencia y de la Comunicación Científica, Centro de Investigación Médica Aplicada, Universidad de Navarra, Pamplona, Spain.
[2] EC3: Evaluación de la Ciencia y de la Comunicación Científica, Departamento de Información y Documentación, Universidad de Granada, Granada, Spain.
[3] Departamento de Física, Universidad de Alcalá, Alcalá de Henares, Spain.

## Abstract

**Purpose:** The aim of this study is to analyze the disciplinary coverage of the Thomson Reuters' Book Citation Index database focusing on publisher presence, impact and specialization.
**Design/Methodology/approach:** We conduct a descriptive study in which we examine coverage by discipline, publisher distribution by field and country of publication, and publisher impact. For this the Thomson Reuters' Subject Categories were aggregated into 15 disciplines.
**Findings:** 30% of the total share of this database belongs to the fields of Humanities and Social Sciences. Most of the disciplines are covered by very few publishers mainly from the UK and USA (75.05% of the books), in fact 33 publishers concentrate 90% of the whole share. Regarding publisher impact, 80.5% of the books and chapters remained uncited. Two serious errors were found in this database. Firstly, the Book Citation Index does not retrieve all citations for books and chapters. Secondly, book citations do not include citations to their chapters.
**Research limitations/implications:** The Book Citation Index is still underdeveloped and has serious limitations which call into caution when using it for bibliometric purposes.
**Practical implications**: The results obtained from this study warn against the use of this database for bibliometric purposes, but opens a new window of opportunities for covering long neglected areas such as Humanities and Social Sciences. The target audience of this study is librarians, bibliometricians, researchers, scientific publishers, prospective authors and evaluation agencies.
**Originality/Value:** There are currently no studies analyzing in depth the coverage of this novel database which covers monographs.

**Keywords:** Book Citation Index; Thomson Reuters; Databases; Citation analysis; Humanities; Social Sciences; Monographs; Publishers

**Article Classification:** Research paper

## Introduction

The launch in October 2010 of Thomson Reuters' newest product for scientific literature the Book Citation Index (hereafter BKCI), may well be a historic milestone in the field of scientometrics or just another failure in the race to develop tools capable of evaluating monographs. Books have long been neglected by bibliometricians (Giménez-Toledo & Román-Román, 2009) mainly due to the lack of data sources and the ways in which they differ from scientific journal literature in terms of their ageing (Glänzel & Schoepflin, 1999; Hicks, 2004; Nederhof, 2006). Although many approaches have been considered before, especially focusing on the fields of the Arts & Humanities as well as the Social Sciences (Torres-Salinas and Moed, 2009; White *et al.*, 2009; Linmans, 2010; Zuccala & van Leeuwen, 2011) where this type of document plays a greater role than in other disciplines (Williams *et al.*, 2009), no definite answer seems to fully convince the research community.

**Corresponding author:**
Nicolás Robinson-García, EC3: Evaluación de la Ciencia y la Comunicación Científica, Facultad de Comunicación y Documentación, Universidad de Granada, 18071, Granada, Spain.
elrobin@ugr.es

In this sense, the launch of this product has an even higher impact in the field, as Thomson Reuters Web of Science, which is highly considered by the scientific community; not only was the first provider of citation indexes but, in their own words, covers only 'top tier international and regional journal literature' (Testa, 2010). What is more, this database has become the main pillar on which many countries have built their whole research policy system. Fully developed by Eugene Garfield in 1972 (Garfield, 2007), the Science Citation Index and the subsequent Social Science Citation Index along with their Journal Citation Reports in which journals are ranked according to the widely known Journal Impact Factor (hereafter JIF) (Garfield & Sher, 1963) have become pivotal within the scientific communication system. Not only has this database dominated the scene as the main source of multidisciplinary scientific literature for nearly fifty years, but its JIF has been established as the legitimate yardstick for measuring prestige among scientific journals. However, it has long been criticized for its incapacity to fully develop a source capable of overcoming the lack of coverage of the Social Sciences and Humanities due to the important role books play in these disciplines (Hicks, 2004). This weakness was in fact, already acknowledge by Garfield himself when he stated:

"From the perspective of the social scientist or humanities scholar, the failure to include monographs as sources in the ISI citation indexes may be a drawback in drawing conclusions about the impact of certain work. [..] Undoubtedly, the creation of a Book Citation Index is a major challenge for the future and would be an expected by-product of the new electronic media with hypertext capability!" (Garfield, 1996)

However, since 2004, with the launch of Scopus — a multidisciplinary database which includes bibliometric data owned by Elsevier, the largest scientific publisher in the world — as well as Google's intrusion in the same year with the emergence of Google Scholar; signified a great threat to Thomson Reuters' Web of Science long monopoly. Soon enough, many studies emerged comparing Web of Science's coverage with that of these new scientific sources (Meho and Yang, 2007) with a special focus on those fields less well-covered by this database (Moed, 2005). In this line of work, Sivertsen & Larsen (2012) analyzed the complete publication output data of Norway finding that, while journals in Humanities and the Social Sciences are quite disperse and influenced by the national factor, book publishing is concentrated in a few publishers. Lacking of fully reliable resources to assess publication types other than journals, Bar-Ilan (2010) performed a microscopic study comparing citation counts to a single book in Scopus, Web of Science and Google Scholar, concluding that the latter offered better results. In a similar paper, Baneyx (2008) compared citation metrics in Humanities and the Social Sciences according to the Web of Science and Google Scholar, arguing that the former is unsuitable for these fields due to the lack of indexed monographs.

Here again, the importance of considering other publication types rises as an unsolved issue. Hicks (2004) comments on the importance of monographs for assessment in the Social Sciences and Humanities arguing that a restriction to journals offers a biased view of the picture, as demonstrated by Glänzel & Schoepflin (1999). In this sense, it is important to mention the interesting project (Martin *et al.*, 2010) developed by different European countries which aims at exploring the possible development of a database that captures the many idiosyncrasies of these fields. But these fields are not the only ones substantially affected by this restriction, even those such as Medicine where journals are the main channel of communication may also be influenced when other publication types are excluded (Lewison, 2004).

## Literature Review

Due to its youth, few studies can be found in the literature referring to the use of the BKCI for evaluation purposes or describing its internal characteristics; coverage, limitations, etc. To date, only two studies have been found. Leydesdorff & Felt (2012) analyze the citation rates of books, book chapters and edited volumes and compare the results offered by the BKCI with those of the other citation indexes. On the other hand, Torres-Salinas *et al.* (2012) propose the development of a 'Book Publishers Citation Reports' and analyze the strengths and weaknesses of such attempt in the Social Sciences and Humanities fields. These types of seminal studies dissecting the coverage, caveats and limitations are considered of great regard as they serve to validate the accuracy and reliability of sources. The launch of Scopus (Elsevier) and Google Scholar and Books have lead to the emergence of many studies analyzing these alternative databases and their advantages and weaknesses when compared with the Thomson Reuters citation indexes (e.g., Kulkarni et al, 2007; Kousha, Thelwall & Rezaie, 2011). But still, books and book chapters seem to lack of widely accepted databases by funding agencies from which to retrieve bibliometric data (Edwards, 2012). Thus, the importance of monographs in the Social Sciences and the Arts & Humanities has lead bibliometricians and research policy managers to search for information resources which would enable them of the necessary tools to assess researchers from these areas.

In this sense, many studies have been conducted analyzing the role of books in research monitoring and alternative citation-enhanced databases which could be used to retrieve bibliometric data. For instance, Kousha & Thelwall (2009) use Google Books Search to compare citations from books with citations from journal articles concluding that the former could complement and even substitute the latter in the fields of the Social Sciences and the Arts & Humanities. In a further study, Kousha, Thelwall & Rezaie (2011) compare the citations to sample of books submitted to the U.K. Research Assessment Exercise retrieved from Google Scholar and Books with those from Scopus. Also, Chen (2012) deepens on the coverage of Google Books comparing it with the OCLC WorldCat. These studies position Google Scholar and Books as possible resources for assessing monographs. However, the launch of the Book Citation Index by Thomson Reuters adds a new alternative to the current landscape, and as such must be analyzed as a previous step to any consideration as an option for bibliometric and evaluative purposes as well as a new information resource to add to library collections.

**Research questions and target audience**
Hence, and since the emergence of the BKCI has been received by the community with great expectancy, we here conduct an in-depth study of its coverage and briefly explore the disciplinary distribution by publisher, output and impact. Our main aim is to analyze BKCI coverage by focusing on identifying the main publishers represented. More specifically, we address at the following research questions:
1. How is the BKCI distributed by discipline and country? Which are the main publishers at each level of analysis?
2. What is the profile of the main BKCI publishers in terms of their output and disciplinary focus?
3. What impact do the mainstream publishers make in terms of their citation rates and which publishers can therefore be considered more prestigious and influential?

In this sense, this type of studes in which a new information source is evaluated and analyzed are quite common in the literature because they provide valuable information not just for evaluation or bibliometric purposes, but also for researchers when choosing tools for information retrieval or even academic librarians who must be especially careful and selective when acquiring and subscribing to new databases. Therefore the target audiences of this study are:
- Bibliometricians. As this type of study allows them to consider the data sources they will use when conducting their studies considering their coverage, reliability and functionalities.

- Also, they support their findings when understanding some of the phenomena they analyze in order to interpret the results they obtain.
- Librarians and researchers. The results of this particular study may help them to decide on the adequacy of the BKCI as an answer to their users' demands or as a data source. Indeed, they often ignore the exact content and scope of the databases they are offered to subscribe to. This study may be a good starting point to understand the potential use of this database.
- Scientific publishers. This study specially focuses on offering an image of how they are represented in the BKCI. It is of special interest to them how they are pictured and how they perform when compared with competitors.
- Prospective authors. The present paper offers a general view which will provide them with information over the coverage by fields of the BKCI and the main publishers represented in it.
- Evaluation agencies. If considered as an evaluation tool, as other Thomson Reuters' citation indexes already are, evaluation agencies should be informed of the strengths and weaknesses of the BKCI before including it as a new tool to be used in their research assessment exercises. Also, the BKCI may provide a useful framework from which to learn when is a book highly cited, which are the citation threshold by area and so on.

## Methods

In this paper we conduct an analysis of Thomson Reuters' Book Citation Index for 2005-2012. In May 2012 we downloaded all BKCI records and created a relational database for data processing and to calculate the indicators. Publisher names were normalized as many had variants that differed as a function of the location of their head offices in each country. For instance, Springer uses variants such as Springer-Verlag Wien, Springer-Verlag Tokyo, Springer Publishing Co, among others. The 249 subject categories, to which records from BKCI were assigned, were also restructured into the 15 disciplines proposed by Moed (2005) in a study of Web of Science journal coverage. However, 2% of the total share could not be included as the records were not assigned to any subject category. In Table 1 we describe the indicators used in our study.

Table 1. Set of bibliometrics indicators calculated to characterize scientific publishers in the Book Citation Index

| Indicator | Acronym | Definition |
| --- | --- | --- |
| Number or percentage of books | Book | Records indexed as document type 'book' in the BKCI |
| Number or percentage of book chapters | Book Chapter | Records indexed as document type 'book chapter' in the BKCI |
| Number of publishers | ScPub | Scientific publishers assigned to books and book chapters |
| Concentration Index | GINI | The Gini Index is usually used in the field of Economics. Here it is defined as an indicator for measuring the concentration of publishers covering each discipline and also for measuring publishers' concentration in the 15 fields defined, in order to interpret the level of specialization. It gives values between 0 and 1: 0 indicates no concentration in a publisher or discipline; 1 indicates concentration in one publisher or discipline. For this indicator, we only considered the document type 'book'. |
| Field Normalized Citation Score | FNCI | The indicator FNCI or Crown is a oriented-field normalized citation score average developed at the CWTS at Leiden University. According to the definition given by Lundberg (2007), it is calculated dividing the average number of received citations for a group of publications (CPP in CWTS terminology) with the average number that could be expected for publications of the same type, from the same year, published within the same field (FCSm in CWTS terminology). Dividing the actual citation rate with the expected citation rate CPP/FCSm we obtain the FNCI. In this study FNCI has been used to measure publisher's impact. We have used for FCSm the 249 BKCI subject categories, allowing us to compare publishers from different areas. A value above 1 is a value above the world's average. |
| Activity Index | Activity Index | The share of a publisher's papers in a specific discipline, divided by the database share of papers in that discipline. A value above 1 means greater activity in the discipline than that expected from database distribution. This indicator shows how much publishers specialize in a certain field. |

## Results

This section is structured as follows. Firstly, we briefly introduce a general perspective of the BKCI. Then, we present a subsection for each of the research questions. We analyze the disciplinary coverage of the database and each country's representation by publisher affiliation. Then, we focus on publisher presence, analyzing who contributes most to the BKCI and their disciplinary profile. Finally, we present a brief descriptive analysis of publisher impact in terms of citation data provided by Thomson Reuters. We must point out that the results shown are related to the actual coverage of the BKCI and not to the book market sector. Therefore the results of this section reflect the main coverage characteristics of this database.

The BKCI has a total of 408 700 records divided in 29 618 books (162 with no ISBN) and 379 082 book chapters (2807 with no ISBN), averaging 13.78 book chapters per book. From the total number of books, 8221 have ISSNs. This reflects that some of the records consider as books by Thomson Reuters are actually serials. Considering their publication date, books seem to be homogeneously distributed with ranges between 4% and 6% each year, except for 2011 when the document share was largest (19% of the database). In fact, in 2009, 2010 and 2011 similar numbers of books have been indexed (5720, 5798 and 5517 respectively), which means numbers seem to have stabilized. Regarding book chapters, the distribution per year varies from 4% in 2009 to 9% in 2005 with 2012 being the year when the greatest share of book chapters can be found (21% of the total).

*Coverage of scientific publishers by discipline -, country and language*

Data is distributed within the 249 Web of Science subject categories. In Table 2, we present records within 15 research fields corresponding with those analyzed by Moed [15]. An effort has been made in the selection process to cover those fields considered by Moed to have moderate coverage, that is, the Humanities & Arts and Other Social Sciences. Each of these disciplines has around 30% of the total share, taking into account books or book chapters. In fact, they are the fields where more publishers have been indexed (143 in Humanities and 114 in Other Social Sciences). The field with the third highest share of the total database is Engineering, with around 12% and 83 different publishers.

Table 2. Book Citation Index coverage and main publishers per scientific fields for the 2005-2012 time period

| Scientific Field | Nr Book | Nr Book Chapter | % Book | % Book Chapter | Nr ScPub | GINI | TOP 3 Publisher in the Field according Nr Book contributing to the field |
|---|---|---|---|---|---|---|---|
| OTHER SOCIAL SCIENCES | 8045 | 96744 | 27,9% | 26,2% | 114 | 0.882 | Palgrave 25%, Routledge 25% and Springer 12% |
| HUMANITIES & ARTS | 8333 | 96007 | 28,9% | 26,0% | 143 | 0.900 | Palgrave 26%, Routledge 15% and Springer 10% |
| ENGINEERING | 3446 | 42895 | 11,9% | 11,6% | 83 | 0.901 | Springer 68%, Nova Science 7% and CRC 5% |
| ECONOMICS | 2843 | 36124 | 9,8% | 9,8% | 64 | 0.859 | Springer 22%, Palgrave 20% and Routlegde 17% |
| CLINICAL MEDICINE | 1633 | 32325 | 5,7% | 8,8% | 59 | 0.851 | Springer 47%, Nova Science 13% and Humana Press 10% |
| BIOLOGICAL SCIENCES - HUMANS | 1544 | 22515 | 5,3% | 6,1% | 68 | 0.826 | Springer 44%, Elsevier 10% and Nova Science 7% |
| MOLECULAR BIOLOGY AND BIOCHEMISTRY | 1119 | 19098 | 3,9% | 5,2% | 38 | 0.815 | Elsevier 25%, Springer 25% and Humana Press 24% |
| BIOLOGICAL SCIENCES - ANIMALS & PLANTS | 1089 | 16904 | 3,8% | 4,6% | 70 | 0.783 | Springer 26%, Cabi Pub.13% and Nova Science 9% |
| MATHEMATHICS | 1522 | 16718 | 5,30 | 4,5% | 55 | 0.858 | Springer 58%, Birkhauser 7% and Siam 4% |
| GEOSCIENCES | 1189 | 16139 | 4,10 | 4,4% | 78 | 0.810 | Springer 38%, Nova Science 10% and Geological Soc 7% |
| APPLIED PHYSICS AND CHEMISTRY | 1087 | 13853 | 3,80 | 3,8% | 31 | 0.823 | Springer 52%, Woodhead 15% and Transtech 15% |
| OTHER SOCIAL SCIENCES - MEDICINE & HEALTH | 671 | 9074 | 2,30 | 2,5% | 51 | 0.707 | Springer 18%, Nova Science 16% and Palgrave 12% |
| CHEMISTRY | 766 | 8016 | 2,7% | 2,2% | 29 | 0.821 | Springer 47%, Royal Society Chem. 17% and Nova Science 8% |
| PHYSICS & ASTRONOMY | 501 | 5740 | 1,7% | 1,6% | 32 | 0.788 | Springer 60%, Nova Science 8% and Annual Reviews 6% |

When focusing on the distribution of publishers by discipline through their concentration index — which shows the distribution of books within fields for a certain publisher — we observe that most of these are covered by few publishers. Other Social Sciences-Medicine & Health presents a higher distribution of records per publisher with a Gini Index of 0.707, while most records in Engineering and Humanities & Arts are distributed over very few publishers presenting a Gini Index of 0.901 and 0.900, respectively. If we observe the three top publishers per area, we find that only for Biological Sciences-Animals & Plants Other Social Sciences-Medicine & Health the top three publishers cover less than half of the content. For the rest of the areas, most records are concentrated in only three publishers. Applied Physics and Chemistry followed by Engineering are the two fields where these publishers concentrate a higher share of the total with 82% and 80%, respectively. Springer seems to be the most multidisciplinary publisher as it is present in all fields, followed by Nova Science, which is one of the top 3 in eight of the fifteen disciplines.

Figure 1. Distribution of the Book Citation Index according to each publisher's affiliation country and language. 2005-2012 time period

**Figure 1A. Distribution of the Book Citation Index according to each publisher's affiliation country**

| Country | Nr Publ. | % Books | Country | Nr Publ. | % Books |
|---|---|---|---|---|---|
| 1. UNITED STATES | 134 | 35,82% | 12. CZECH REPUBLIC | 6 | 0,02% |
| 2. ENGLAND | 50 | 39,23% | 13. BRAZIL | 5 | 0,03% |
| 3. GERMANY | 25 | 13,68% | 14. PEOPLES R CHINA | 4 | 0,05% |
| 4. NETHERLANDS | 12 | 5,95% | 15. SPAIN | 4 | 0,03% |
| 5. ITALY | 8 | 0,15% | 16. AUSTRIA | 3 | 0,86% |
| 6. POLAND | 8 | 0,07% | 17. SINGAPORE | 3 | 0,58% |
| 7. SWITZERLAND | 7 | 1,44% | 18. HUNGARY | 3 | 0,03% |
| 8. CANADA | 7 | 0,60% | 19. SCOTLAND | 2 | 0,25% |
| 9. FRANCE | 7 | 0,31% | 20. JAPAN | 2 | 0,06% |
| 10. FINLAND | 6 | 0,34% | 21. SLOVENIA | 2 | 0,03% |
| 11. AUSTRALIA | 6 | 0,31% | 22. ROMANIA | 2 | 0,01% |

**Figure 1B. Distribution of the Book Citation Index according to language**

| Language | Nr Books | % Books | Language | Nr Books | % Books |
|---|---|---|---|---|---|
| 1. ENGLISH | 28826 | 96,73% | 10. RUSSIAN | 4 | 0,01% |
| 2. GERMAN | 727 | 2,44% | 11. LATIN | 4 | 0,01% |
| 3. FRENCH | 183 | 0,61% | 12. POLISH | 5 | 0,02% |
| 4. SPANISH | 35 | 0,11% | 13. RUSSIAN | 4 | 0,01% |
| 5. FINNISH | 34 | 0,12% | 14. ESTONIAN | 3 | 0,01% |
| 6. ITALIAN | 31 | 0,10% | 15. SWEDISH | 3 | 0,01% |
| 7. PORTUGUESE | 14 | 0,05% | 16. GREEK | 3 | 0,01% |
| 8. SLOVENIAN | 5 | 0,02% | 17. GEORGIAN | 2 | 0,01% |
| 9. DUTCH | 5 | 0,02% | 18. CZECH | 2 | 0,01% |

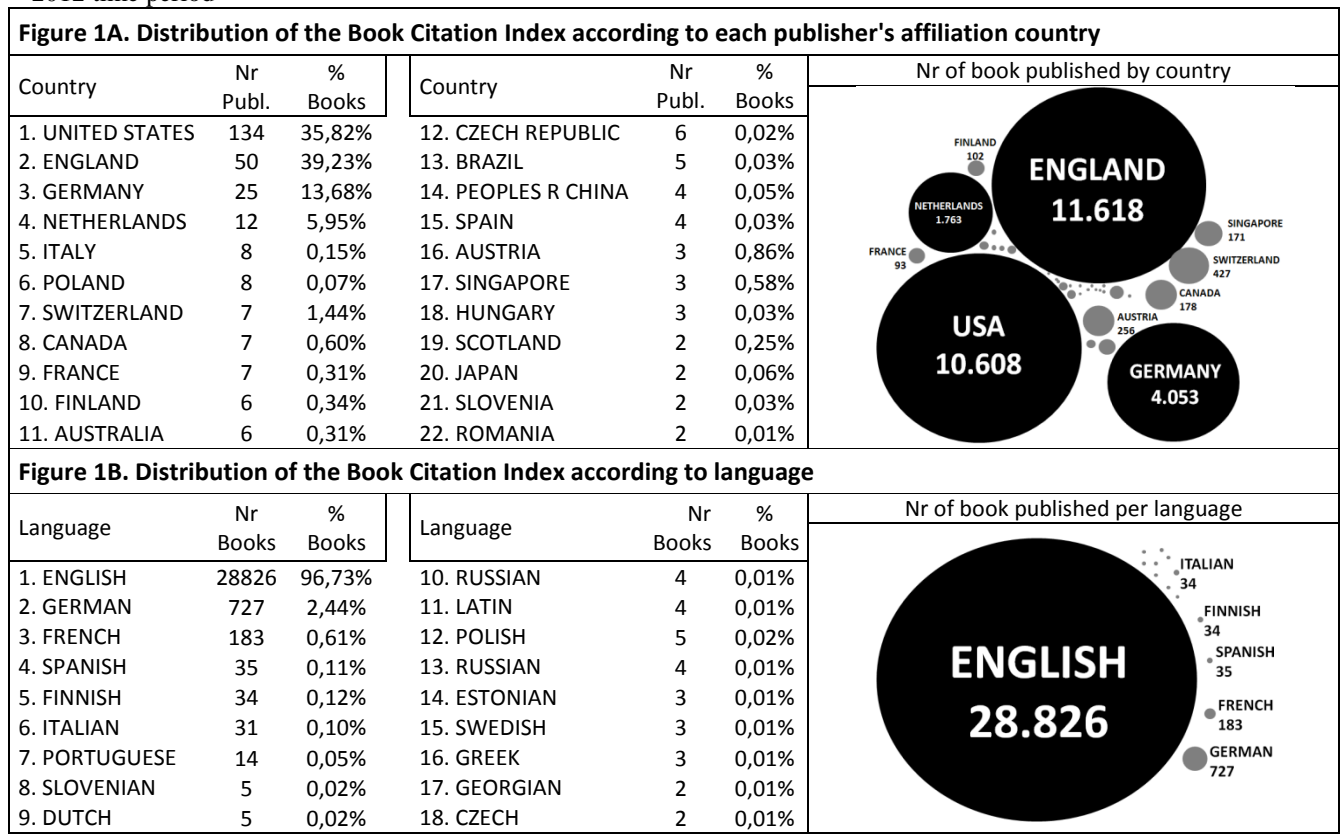

Regarding publisher affiliation (Figure 1A), we must emphasize that countries are assigned according to where the book was published. That means, for instance, that books from Springer can be published in several countries. We find a strong bias towards English-speaking countries. In fact, the United States and England represent 56% of publishers (194) and 75.05% of the whole database taking into account books as publication type. However, when analyzing the distribution by region, we observe that European countries represent 62.4% of records while North America represents 36.42%, Asia and Oceania 1% and South America is only represented by Brazil with 0.03%. When only focusing on publishers and not records, we see that the US is by far the country with more BKCI indexed publishers, with three times more than second-placed England, with 50. The third-ranked country is Germany with 25 publishers, half as many as England. Regarding the distribution by languages (Figure 1B), the database is greatly biased towards the English language as many publishers affiliated to the Netherlands or Germany publish in English and not in the language of the publication country (i.e., Elsevier). This can be confirmed when observing the distribution by languages where around 96% of the whole database are books in English followed by German books (2.44%). The rest of the languages do not even reach 1% of the total share of the BKCI.

*Publisher production and specialization*

In Figure 2 we present five BKCI publishers according to their concentration index (GINI), showing whether or not they have a specialized or a multidisciplinary focus. These publishers account for more than 60% of the BKCI in terms of books or book chapters. Therefore, they can be considered the mainstream publishers in the database. As suggested by Table 1, we find that Springer and Nova Science have multidisciplinary profiles with values of 0.380 and 0.294, respectively. On the other hand, Palgrave with a GINI score of 0.825 and Routledge with 0.820 are publishers that specialized in highly specific disciplines as their acute curves show. Finally, Cambridge University Press is somewhere in between with a GINI of 0.684.

Figure 2 Level of specialization for the top five publishers in the Book Citation Index for the 2005-2012 time period

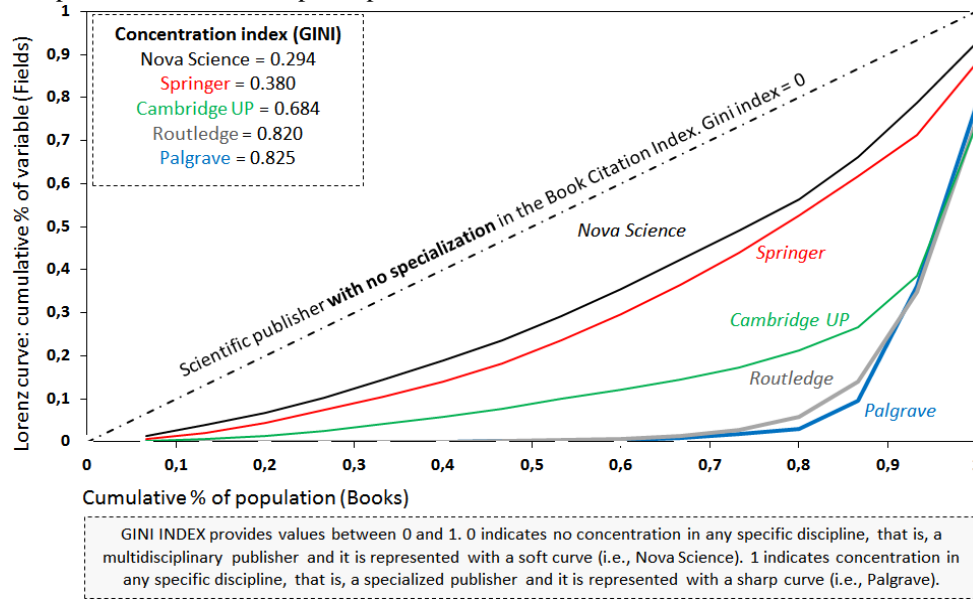

In Table 3 we present the top 33 BKCI publishers ranked by the number of books published. They represent 90% of total coverage and each has at least 100 books indexed in the database. By far the most prolific publishers in terms of books are Springer (7783), Palgrave (4164) and Routledge (3369). However, if we analyze the disciplines they mainly specialize in, we observe different patterns. While Nova Science, Springer, Blackwell Science and Annual Reviews have multidisciplinary profiles that never surpass 0.4 of their concentration index, 18 of the top 33 publishers have highly specialized profiles which all pass the 0.8 concentration index barrier. Of these, only six present a profile focused on Health Sciences while the other twelve specialize in Social Sciences and Humanities. Humana Press for instance has an activity index of 14.02 for Molecular Biology and Chemistry, but the highest rate is found that of Woodhead Publishing with an activity index of 30.64. Palgrave and Routledge, the two second-placed mainstream publishers in BKCI are highly specialized in the Social Sciences and Humanities.

Table 3. Mainstream Book Citation Index publishers and their main specializations for 2005-2012

| Publisher* | Main Country | Nr Book | % Book | Nr Book Chapter | % Book Chapter | GINI | Main specialization fields - Activity Index** | | | |
|---|---|---|---|---|---|---|---|---|---|---|
| SPRINGER | Netherlands | 7783 | 26.28% | 101975 | 26.90% | 0.380 | ENG 2.45 | MAT 2.19 | CHE 1.77 | MED CL 1.84 |
| PALGRAVE | England | 4164 | 14.06% | 45350 | 11.96% | 0.825 | SOC 1.81 | HUM 1.80 | ECO 1.40 | SOC MED 0.83 |
| ROUTLEDGE | England | 3399 | 11.48% | 42017 | 11.08% | 0.820 | SOC -2.15 | PSY -1.68 | ECO -1.42 | HUM -1.25 |
| NOVA SCIENCE | United States | 1318 | 4.45% | 14168 | 3.74% | 0.294 | SOC MED 3.43 | PHY AP 2.85 | MED CL 2.78 | PSY 2.18 |
| CAMBRIDGE UNIV PRESS | England | 1095 | 3.70% | 14566 | 3.84% | 0.684 | HUM 2.04 | PSY 1.64 | SOC 0.74 | SOC MED -0.71 |
| WALTER DE GRUYTER & CO | Germany | 962 | 3.25% | 13429 | 3.54% | 0.872 | HUM 3.07 | MAT 0.69 | SOC 0.38 | BIO HU 0.31 |
| ELSEVIER | Netherlands | 733 | 2.47% | 8247 | 2.18% | 0.621 | BIO MO 10.16 | PHY AP 4.65 | BIO HU 4.13 | CHE 2.83 |
| PRINCETON UNIV PRESS | United States | 715 | 2.41% | 7857 | 2.07% | 0.763 | ECO 1.89 | HUM 1.85 | SOC 1.62 | MAT 1.18 |
| EDWARD ELGAR PUBLISHING | England | 616 | 2.08% | 8055 | 2.12% | 0.818 | ECO 6.89 | GEO 1.36 | SOC 1.20 | SOC MED 0.49 |
| UNIV CALIFORNIA PRESS | United States | 602 | 2.03% | 7288 | 1.92% | 0.801 | BIO AP 2.32 | HUM 1.85 | SOC 1.53 | SO ME 1.20 |
| HUMANA PRESS INC | United States | 543 | 1.83% | 11883 | 3.13% | 0.821 | BIO MO 14.02 | ME CL 5.81 | BIO HU 2.96 | SO ME 1.38 |
| UNIV PENNSYLVANIA PRESS | United States | 465 | 1.57% | 4025 | 1.06% | 0.869 | HUM 2.71 | SO 1.37 | ECO 0.37 | SO ME 0.28 |
| CRC PRESS-TAYLOR & FRANCIS | United States | 373 | 1.26% | 5801 | 1.53% | 0.506 | PHY AP 6.33 | CHE 5.30 | ENG 3.71 | BIO AP 2.44 |
| EMERALD GROUP | England | 319 | 1.08% | 4152 | 1.10% | 0.824 | ECO 4.61 | SO ME 2.04 | SO 1.90 | HUM 0.28 |
| WOODHEAD PUBL LTD | England | 309 | 1.04% | 5566 | 1.47% | 0.790 | PHY AP 30.64 | BIO AP 7.05 | ENG 1.64 | BIO MO 0.26 |
| INFORMATION AGE | England | 287 | 0.97% | 3641 | 0.96% | 0.848 | SOC 2.97 | PSY 1.68 | ECO 1.32 | HUM 0.43 |
| BLACKWELL SCIENCE PUBL | England | 281 | 0.95% | 4813 | 1.27% | 0.397 | CHE 5.83 | SO ME 3.14 | PHY AP 2.72 | BIO HU 2.25 |
| ANNUAL REVIEWS | United States | 252 | 0.85% | 5738 | 1.51% | 0.368 | BIO HU 3.80 | BIO AP 3.60 | PHY AP 3.42 | BIO MO 2.30 |
| CABI PUBLISHING-C A B INT | England | 236 | 0.80% | 4075 | 1.07% | 0.715 | BIO AP 15.59 | GEO 3.65 | BIO MO 2.33 | SOC 1.02 |
| M I T PRESS | England | 210 | 0,71% | 2796 | 0,74% | 0,641 | PSY 3,22 | GEO 2,61 | SO MED 1,47 | ECO 1,34 |
| ROYAL SOC CHEMISTRY | England | 207 | 0,70% | 2553 | 0,67% | 0,669 | CHE 23,98 | BIO MO 3,63 | ENG 1,67 | BIO HU 1,54 |
| UNIV NORTH CAROLINA PRESS | United States | 199 | 0,67% | 2104 | 0,56% | 0,877 | HUM 2,97 | SOC 0,94 | SO MED 0,92 | GEO 0,26 |
| CHANDOS PUBL | England | 193 | 0,65% | 2118 | 0,56% | 0,884 | SOC 2,70 | ECO 0,96 | HUM 0,62 | ENG -0,04 |
| E J BRILL | Netherlands | 188 | 0,63% | 2242 | 0,59% | 0,867 | HUM 2,75 | BIO AP 1,23 | SOC 0,78 | GEO 0,32 |
| AUSTRALIAN NATL UNIV | Australia | 183 | 0,62% | 2463 | 0,65% | 0,792 | GEO 2,29 | SOC 1,95 | HUM -1,75 | ECO 1,30 |
| BIRKHAUSER VERLAG AG | Switzerland | 143 | 0,48% | 1448 | 0,38% | 0,807 | MAT 13,14 | MED CL 0,99 | PHY AP 0,81 | BIO HU 0,65 |
| ARTECH HOUSE | United States | 124 | 0,42% | 1492 | 0,39% | 0,847 | PHY AP 10,23 | ENG 7,50 | GEO 1,18 | MED CL 1,00 |
| MANEY PUBLISHING | England | 117 | 0,40% | 906 | 0,24% | 0,932 | HUM 3,47 | SOC 0,03 | --- | --- |
| INTELLECT LTD | England | 114 | 0,38% | 1401 | 0,37% | 0,881 | SOC 2,54 | HUM 1,50 | ECO 0,09 | --- |
| EDITIONS RODOPI B V | Netherlands | 104 | 0,35% | 1628 | 0,43% | 0,926 | HUM 3,27 | SOC 0,21 | --- | --- |
| KARGER | Switzerland | 103 | 0,35% | 1567 | 0,41% | 0,839 | MED CL 14,11 | BIO HU 3,98 | PSY 2,98 | SO MED 2,29 |
| IOS PRESS | Netherlands | 102 | 0,34% | 2434 | 0,64% | 0,737 | ENG 5,34 | SOC ME 2,53 | ECO 1,49 | SOC 1,30 |
| WORLD SCIENTIFIC | Singapore | 102 | 0,34% | 1304 | 0,34% | 0,548 | PHY AP 7,73 | MAT 6,46 | PHY AS -2,47 | ENG 2,42 |

* Publishers have at least 100 books indexed and they concentrate 90% of the BKCI total output regarding books as well as book chapters
** **PHY AP** - Applied Physics and Chemistry | **BIO AP** - Biological Sciences - Animals & Plants | **BIO HU** - Biological Sciences - Humans | **CHE** - Chemistry | **MED CL** - Clinical Medicine | **ECO** - Economics | **ENG** - Engineering | **GEO** - Geosciences | **HUM** - Humanities & arts | **MAT** - Mathematics | **BIO MO** - Molecular Biology and Biochemistry | **SOC** - Other Social Sciences | **SOC ME** - Others Social Sciences - Medicine & Health | **PHY AS** - Physics & Astronomy | **PSY** - Psychology & Psychiatry

*Impact*

The 408 700 records indexed by the BKCI have received a total of 680 680 citations, this means an average of 1.67 citations per record. Citation counts to books or book chapters include citations from BKCI but also from all the rest Web of Science databases as for example AH&CI, SSCI or SCI. However, as described in Table 4, 80.5% of the total publications have received no citations and 1% of the total output has more than 16 citations per publication. Publications with a range between 0 and 5 citations represent 95.1% of the total database.

Table 4. Number of publications per range of Book Citation Index citations for 2005-2012

| TIMES CITED | RECORDS | % RECORDS |
|---|---|---|
| 0 | 329 172 | 80.5% |
| 1-5 | 59 855 | 14.6% |
| 6-10 | 8232 | 2.0% |
| 11-15 | 3312 | 0.8% |
| 16-20 | 1870 | 0.5% |
| 21-25 | 1219 | 0.3% |
| >25 | 5040 | 1.2 |
| TOTAL | 408 700 | 98.8 % |

However, the most interesting thing is that one single publisher, Annual Reviews, represents 41% of the total citation count. Actually, records from this publisher have an average of 46.45 citations per *document*. This means that if we exclude this publisher, the average citation rate drops to 0.99 per document. When looking at the nature of the books published by Annual Reviews, we find most are actually periodicals as not only are they ordered by volumes but they also have an ISSN assigned, greatly distorting the database as the publisher with the second highest average of citations per document, Oxford University Press, has 12.35. In fact, most of the records of this publisher are review articles. As pointed out by Archambault and Larivière (2009), journals with the highest Journal Impact Factor are usually review journals, simply because review articles are more frequently cited than regular articles a factor that contribute to the anomalous performance of Annual Reviews. This distortion is so great that it forced us to exclude it in order to represent publishers according to their impact and concentration index as in Figure 3.

Figure 3. Field Normalized Citation Score and Concentration Index for publishers according to the Book Citation Index for 2005-2012

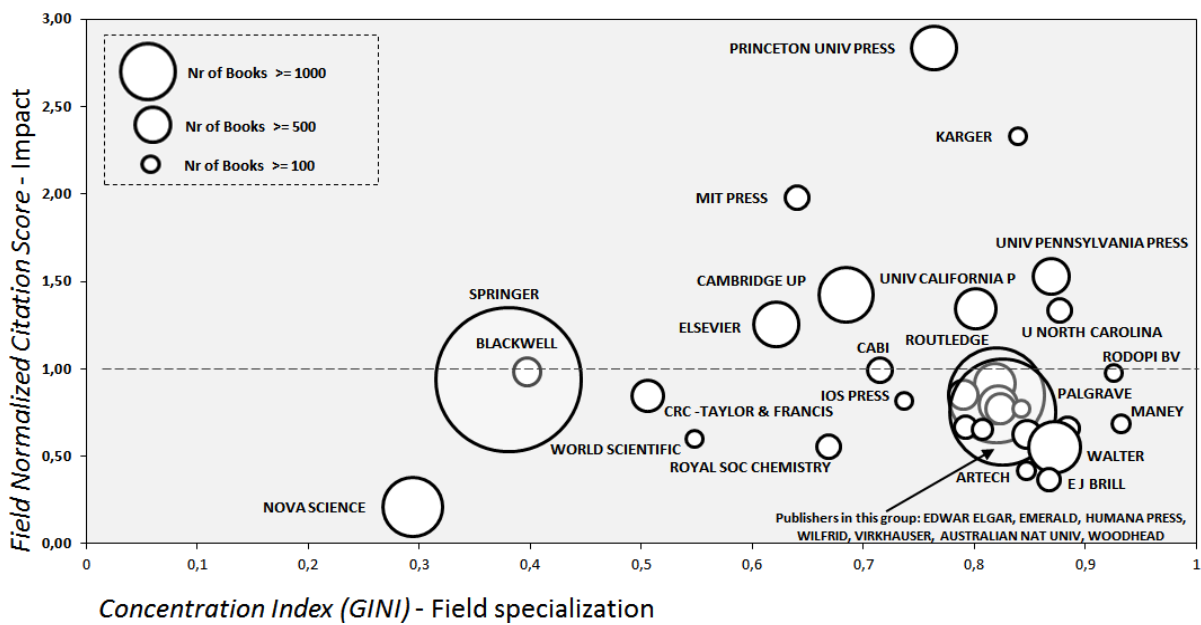

Figure 3 represents publishers according to their Field Normalized Citation Impact, Field Concentration Index (GINI) and number of books. We observe how most publishers are grouped on the right of the graph, showing that most of them display a disciplinary focus. Prestigious university publishers have higher impact (e.g., Princeton University Press, MIT Press, Cambridge

UP), while multidisciplinary publishers tend to have lower impact (e.g., Nova Science, Blackwell, Springer). Also, we observe that Springer and Palgrave and Routledge, which can be considered three of the main scientific publishers in terms of number of books indexed in the BKCI, show lower values than expected, probably due to the high volume of books they generate, but also because their book chapters are little cited. Finally, it is interesting to note that some clusters emerge from this representation. Firstly, that of university publishers, highly specialized and with a high impact (i.e., Princeton, MIT Press, Cambridge and University of California); secondly, that of a second group of highly specialized publishers with low impact (Humana Press, Palgrave, Routledge and Emerald, for instance).

Table 5. Some problems detected in the Book Citation Index citation count for 2005-2012

| PROBLEM A) BKCI does not retrieve all citations for books and book chapters. When using the *Cited Reference Search* option from Web of Science, more citations are identified rather than those shown by BKCI. This is a data processing problem. | | | |
| --- | --- | --- | --- |
| EXAMPLES: | Nr of Citation in the BKCI (Books) | Nr of Citation in the BKCI (Book Chapters) | Nr of Citations using 'Cited Ref. Search' |
| ● Moed, HF. *Citation Analysis in Research Evaluation*. SPRINGER, 2005 | 7 | 12 | 298 |
| ● Griewank, A; Walther, A. *Evaluating Derivatives: Principles and Techniques of Algorithmic Differentiation* (2nd ed.). SIAM, 2008 | 51 | 0 | 305 |
| ● Dill, KA; Bromberg, S. *Molecular Driving Forces: Statistical Thermodynamics in Biology, Chemistry, Physics, and Nanoscience*. GARLAND SCIENCE, 2011 | 4 | 0 | 186 |
| ● Forrester, PJ. *Log-Gases and Random Matrices*. PRINCETON UNIVERSITY PRESS, 2010 | 52 | 0 | 174 |
| PROBLEM B) Book citations do not include their book chapters citations. This is a conceptual problem. | | | |
| EXAMPLES: | | Nr of Citation in the BKCI (Book) | Nr of Citation in the BKCI (Books Chapters) |
| ● Hancock, GR; Mueller, RO. *Quantitative Methods in Education and the Behavioral Sciences-Issues Research and Teaching*. INFORMATION AGE PUBLISHING-IAP. CHARLOTTE, 2006 | | 25 | 267 |
| ● Wood, J; Dupont, B. *Democracy, Society and the Governance of Security*. CAMBRIDGE UNIV PRESS. 2006 | | 25 | 94 |
| ● Briggs, WR; Spudich, JL. *Handbook of Photosensory Receptors*. BLACKWELL SCIENCE PUBL, OXFORD. 2005 | | 6 | 260 |
| ● Lau, WKM; Waliser, DE. *Intraseasonal Variability in the Atmosphere-Ocean Climate System*. SPRINGER-PRAXIS BOOKS IN ENVIRONMENTAL SCIENCES, 2005 | | 69 | 278 |

But we must call for caution when interpreting these results. As with every new, on-going project, we have detected errors on the citation count which may put into question some of the BKCI impact data. In Table 5, we present some examples of the two main mistakes we have found: a data processing problem and a conceptual problem. The first has to do with the data processing as some publications have considerably fewer citations in BKCI than can be seen when searching the book using the Cited Reference Search option. The case of Henk Moed's "Citation Analysis in Research Evaluation" is of special significance. While the BKCI retrieves 7 citations to the book and 12 to book chapters, we have retrieved nearly 300 citations from the Cited Reference Search Option. In fact, if we look further at possible causes we see how the BKCI identifies the book title as the source and therefore does not match most of the citations correctly. But there are also some problems when processing different editions of a single book, for instance, Griewank & Walther's book which only registers citations to their 2nd edition, from 2005. The second error has to do with a conceptual issue. As the BKCI indexes books and book chapters in separate records, it also considers their citations as separate. Therefore a book may have fewer citations than the sum of the citations received by its book chapters, an especially worrying issue when analyzing single-author books. Although this decision may be understandable for edited volumes it seems problematic particularly from a user's viewpoint that would usually be searching for the overall impact of a monograph. Indeed, although it is a multi-authored book, we find that the "Handbook of Photosensory Receptors" by Briggs and Spudich has received 6 citations according to BKCI, but its book chapters have received a total of 260 citations, giving a misleading idea of the book's impact.

**Discussion**

In this paper we analyze BKCI coverage in terms of publishers indexed by country and discipline as well as their impact. For this purpose we posed four research questions regarding the main publishers present in this database, their distribution by country and by discipline, the mainstream publishers' disciplinary focus, and their impact.

We observe that in the BKCI 33 scientific publishers concentrate 90% of the whole share of the database. Indeed three publishers — Springer, Palgrave and Routledge — represent more than half of the books indexed in the BKCI. Interestingly, Springer and Nova Science, the first and fourth most prolific publishers are these with the most multidisciplinary focus, while Palgrave and Routledge are highly specialized in the Social Sciences and Humanities. This shows Thomson Reuters' efforts to overcome their long-attributed weakness in failing to adequately cover these two research fields in the citation indexes. However, along with publication type, Thomson Reuters has been also criticized for its heavy bias towards English-speaking publications (Van Leeuwen *et al.*, 2001), an issue that especially affects the fields of Humanities and Social Sciences where literature is more disperse and the regional factor has a stronger influence (Archambault and Larivière, 2010; Hicks and Wang, 2011). This is has not been achieved by this product, as England and the US alone represent 75.05% of the total share by publication country and English language represents 96% of the database, showing a strong bias towards the Anglo-Saxon world and neglecting the research output of other countries and linguistic communities such as German, French or Spanish languages. In fact, although they acknowledge the importance of other languages rather than English, they emphasize their preference for this language stating that "English language full text is highly desirable, but books with full text in a language other than English are also considered" (Testa, 2010). Also, the great concentration of publishers for these areas (0.882 in Other Social Sciences and 0.900 in Humanities & Arts) can be controversial, as many of these disciplines are characterized by their controversies between different trends of thought and paradigms.

An important issue that must be addressed after analyzing the BKCI is if the coverage of the BKCI represents the book market sector. Contrarily with journals where we can check with Ulrich's Periodicals Directory, one of the main problems we encounter when analyzing this issue is the lack of exhaustive information resources which would allow us to know with precision what does the BKCI represents from the world's share. However, if we consult the statistics of specific countries we observe that the coverage by languages and countries is still very poor. For instance, in Table 6 we show data for the United States and Spain retrieved from the New Book Titles and Editions, 2002-2011 and Estadística de la Producción Editorial 2011. As observed, the US book market sector is three times bigger thant the Spanish one, proportions which are not reflected in the BKCI and which lead us to think that the database is strongly biased towards certain countries and languages.

Table 6. Number of books published in United States and Spain in 2011 according its national statistics

| Discipline | Nr of Books United States | Nr of Books Spain |
|---|---|---|
| Social Sciences | 31633 | 13595 |
| Applied Sciences | 24692 | 10194 |
| Science | 18499 | 2911 |
| **Total** | **74824** | **26700** |

Probably, the most interesting issue to discuss from our study has to do with analyzing publisher impact and how the BKCI processes citation counts. The low citation rates found are quite surprising, especially when the company emphasizes its commitment to indexing the most relevant literature (Testa, 2010). But the uncitedness rate does not seem to differ much to that described by Schwartz (1997) who reports levels of uncitedness of 75% and 92% for Social Science and Humanities respectively. The uncitedness issue is a long-discussed research front in bibliometrics, first mentioned by Price (1965) who stated that 10% of scientific literature never gets cited. In this sense, because the BKCI only covers books dating from 2005 and considering that books and monographs are cited at a slower rate than articles, this high level should probably not be considered suspicious. A wider citation window would probably be needed to analyze this phenomenon. Also, the citation rate of monographs differs from that of journals (Hammarfelt, 2011), calling for caution when interpreting citation counts for each publication type.

But the problems indicated in the citation count should be taken into account and solved. In this sense, those derived from technical issues will probably be corrected soon but others of a more conceptual nature such as the distinction between book citations and book chapter citations, or those regarding the processing of different editions of a single book should be analyzed in depth. Indeed, the inclusion of 40 journals published by Annual Reviews (Palo Alto, US) distorts the nature of the database in terms of citations as its publications should not strictly be considered as books especially if we take into account that 35 of the journals published by Annual Reviews are indexed in the Journal Citation Reports and have Impact Factor.

## Conclusion

The launch of the BKCI by Thomson Reuters means great news to information scientists and bibliometricians, as it provides a new resource for the analysis and development of tools to complete the research evaluation picture. As long-neglected publication types in citation indexes due to their nature, books have become an important flaw in bibliometric studies because they are an important channel of communication in the fields of Humanities and the Social Sciences. Their inclusion may improve the coverage of these fields in the Thomson Reuters' Web of Science. In this study, we have analyzed the product in order to assess its validity from the viewpoint of publisher presence and impact. Indeed, the Humanities and Social Sciences represent more than half of the total database. But the strong presence of English-speaking countries in the BKCI shows a heavy bias towards these countries, demonstrating that the same limitations that have already been denounced for the other citation indexes (Van Leeuwen *et al.*, 2001), also occur here.

Another problem with the coverage of the BKCI is that it includes serials that are actually part of the other journal citation indexes (Science Citation Index, Social Science Citation Index and Arts & Humanities Citation Index) and therefore, are not new inclusions. We have previously mentioned the case of Annual Reviews that clearly should not be part of the BKCI. In relation to this case if Thomson-Reuters considers Annual Reviews as books, then others similar serials should have been included such as the Annual Review of Information Science and Technology (ARIST) that also has ISBN and ISSN. In this sense, Thomson-Reuters lack of a clear policy to distinguish a journal from a book.

With regards to impact, many limitations have been found in the citation count. Although some of them are data-processing errors and should be easily corrected, others are of a conceptual nature. As previously discussed, the issue regarding different editions of a single book, or the differences in count between books and book chapters may well be serious flaws in this database. However, as we have only included records from 2005 in our analysis, more time may well be needed before analysis of impact data and book citation curves is satisfactory. Also further analyses are needed in order to assess on the content of the BKCI and its correspondence with the publishing sector similar to that performed by Chen (2012) regarding Google Books.

**Acknowledgments**

Thanks are due to the three anonymous referees for their constructive suggestions. Thanks are due to Bryan J. Robinson Fryer for reviewing the English version of this paper. This research was sponsored by the Ministerio de Economía y Competitividad under a grant HAR2011-30383-C02-02. Nicolás Robinson-García is currently supported by a FPU grant from the Spanish Ministerio de Economía y Competitividad.